# How a charge conserving alternative to Maxwell's displacement current entails a Darwin-like approximation to the solutions of Maxwell's equations

Alan M Wolsky[1,2 a)b)]
[1] Argonne National Laboratory, 9700 South Cass Ave., Argonne IL 60439
[2] 5461 Hillcrest Ave., Downers Grove, IL 60515

[a)] E- mail: AWolsky@ANL.gov
[b)] E- mail: AWolsky@ATT.net

**Abstract**. Though sufficient for local conservation of charge, Maxwell's displacement current is not necessary. An alternative to the Ampere-Maxwell equation is exhibited and the alternative's electric and magnetic fields and scalar and vector potentials are expressed in terms of the charge and current densities. The magnetic field is shown to satisfy the Biot-Savart Law. The electric field is shown to be the sum of the gradient of a scalar potential and the time derivative of a vector potential which is different from but just as tractable as the simplest vector potential that yields the Biot-Savart Law The alternative describes a theory in which action is instantaneous and so may provide a good approximation to Maxwell's equations where and when the finite speed of light can be neglected. The result is reminiscent of the Darwin approximation which arose from the study classical charged point particles to order $(v/c)^2$ in the Lagrangian. Unlike Darwin, this approach does not depend on the constitution of the electric current. Instead, this approach grows from a straightforward revision of the Ampere Equation that enforces the local conservation of charge.

I.   Introduction

Though sufficient for local conservation of charge, Maxwell's addition of his displacement current to Ampere's Law is not necessary. Other additions could have been made with the same effect, as described elsewhere.[1] Here we show that one such alternative to Maxwell's displacement current entails an expression for the magnetic field that is given by Biot-Savart's Law. The same alternative entails an expression for the electric field that is reminiscent of Darwin's approximation of the solution of Maxwell's Equations, an approximation meant to be used when the sources are relatively slow moving classical point charges.[2] Unlike Darwin's Approximation, our approach does not rely on the constitutive equation for the current. Rather, we present a macroscopic theory that is characterized by its alternative to Maxwell's displacement current.

The rest of this paper is organized as follows. Section II describes the alternative to the displacement current. The magnetic field and vector potential that derive from the alternative are discussed in Section III. Section IV is devoted to the concomitant electric field. Its calculation in terms of its sources, charge and current density, appears difficult but after the work described herein, the final result is not qualitatively more difficult to evaluate than the Biot-Savart Law. Section V summarizes and concludes.

A preliminary remark may be helpful. When discussing the solutions to Maxwell's equations, attention is focused on solutions that describe fields that tend toward zero as the point of observation recedes farther and farther from the fields' sources. Further, the fields' sources, charges and currents, are assumed to lie in a bounded region or to diminish fast enough at infinity to permit one to neglect surface integrals at infinity. The same assumptions will be made here.

II.  An alternative to Maxwell's displacement current

For the reader's ready reference, this section recalls some observations made in an earlier paper which presents a more thorough (and somewhat more leisurely) discussion of them.[1]

In terms of the fields, the four "pre-Maxwell Equations" are as shown below

$$div[\mathbf{E}] = \rho/\varepsilon_0 \qquad (\text{II.1})$$
$$div[\mathbf{B}] = 0 \qquad (\text{II.2})$$
$$\mathbf{curl}[\mathbf{E}] = -\partial_t \mathbf{B} \qquad (\text{II.3})$$
$$\mathbf{curl}[\mathbf{B}] = \mu_0 \mathbf{J} \qquad (\text{II.4})$$

where $\rho$ denotes the total (free and polarization) charge density and $\mathbf{J}$ denotes the total (free and magnetization) current density. And we must not forget, Eq.II.5, the fifth pre-Maxwell Equation; it expresses the local conservation of charge, a requirement that cannot be compromised.

$$div[\mathbf{J}] = -\partial_t \rho \qquad (\text{II.5})$$

Indeed, Eq.II.5 does more than assert current conservation. It relates the current density that is the source of the static magnetic field to the charge density that is a source of the static electric field.

The first four pre-Maxwell Equations describe situations in which the divergence of the current density, $\mathbf{J}$, is zero, that is to say situations in which the charge density is time independent. Of course, this excludes consideration of very important situations, for example, situations in which a capacitor's plates are connected to an alternating current. Another example is provided by an antenna excited by an oscillating current. Thus one wants to know what happens where and when the charge density is changing. As is widely appreciated, the pre-Maxwell Equations cannot provide an answer. The divergence of the left hand side of Ampere's Equation, Eq.II.4. is always zero and so only currents with zero divergence can appear on the right. Thus we come to the question mentioned in every textbook, "How should Ampere's Equation be revised to describe all situations?" Maxwell's answer was to add the term $\partial_t(\varepsilon_0 \mathbf{E})$ and thereby arrive at what is often now often called the Ampere-Maxwell Equation.

$$\mathbf{curl}[\mathbf{B}] = \mu_0 \left( \mathbf{J} + \partial_t(\varepsilon_0 \mathbf{E}) \right) \tag{II.6}$$

It does the job because we believe in the local conservation of charge, as expressed by Eq.II.5. And so the vanishing divergence of the left-hand side of Eq.II.6 is matched to the vanishing divergence of the right-hand side. Alternatively, one could say that Eq.II.6 entails the local conservation of charge, something we are predisposed to believe. In summary, the Ampere-Maxwell Equation is sufficient to guarantee local charge conservation.

But, is it necessary? Is there a different revision to the Ampere Equation that would achieve the same result? Well yes, there is. That different revision comes to mind when one recalls that the divergence of the gradient is the Laplacian and so one has Eq.II.7.

$$div_x \left[ \nabla_x \frac{1}{4\pi \|\mathbf{x} - \mathbf{y}\|} \right] = \Delta_x \frac{1}{4\pi \|\mathbf{x} - \mathbf{y}\|} = -\delta^3[\mathbf{x} - \mathbf{y}] \tag{II.7}$$

Thus Eq.II.8 presents a different generalization of Ampere's Equation.

$$\mathbf{curl}[\mathbf{B}[t,\mathbf{x}]] = \mu_0 \left( \mathbf{J}[t,\mathbf{x}] - \nabla_x \iiint_{all\ space} d^3 y \frac{\partial_t \rho[t,\mathbf{y}]}{4\pi \|\mathbf{x} - \mathbf{y}\|} \right) \tag{II.8}$$

After taking the divergence of both sides, one confirms that Eq.II.8 entails the local conservation of charge and thus our addition presents an alternative to Maxwell's displacement current. Maxwell called the sum of the conduction current and his displacement current the true current and denoted it by $\mathbf{C}$.[3] We will denote the analogous quantity in Eq.II.8 by $\mathbf{J}^{ATM}$ where ATM stands for Alternative to Maxwell.

**III. How the atm's magnetic field depends on currents**

The unusually attentive reader of either Panofsky and Philips or Jackson, might remember the solution of Eqs.II.2 and II.8. [4,5,6] It is the familiar Law of Biot-Savart.

$$\mathbf{B}^{ATM}[t,\mathbf{r}] = \iiint_{all\ space} \frac{d^3 r_s}{4\pi} \frac{\mu_0 \mathbf{J}[t,\mathbf{r}_s] \times (\mathbf{r} - \mathbf{r}_s)}{\|\mathbf{r} - \mathbf{r}_s\|^3} \tag{III.1}$$

The right hand side of Eq. III.1 can be re-expressed as the curl of a vector, as shown in Eq.II.10.

$$\mathbf{B}^{ATM}[t,\mathbf{r}] = \mathbf{curl}_r \left[ \underbrace{\iiint_{all\ space} \frac{d^3 r_s}{4\pi} \frac{\mu_0 \mathbf{J}[t,\mathbf{r}_s]}{\|\mathbf{r}-\mathbf{r}_s\|}}_{\equiv \mathbf{A}^{ATM}_{B-S}} \right] \qquad (\text{III.2})$$

In this form, the argument of the curl – which will be denoted by $\mathbf{A}^{ATM}_{B-S}$ – brings to mind the vector potential that often appears in discussions of magnetostatics. In that situation, its divergence is zero and so it satisfies the definition of the Coulomb gauge. In general, it does not. Indeed, straightforward manipulation shows that the divergence of the Biot-Savart vector potential $\mathbf{A}^{ATM}_{B-S}$ is proportional to the divergence of the current density, $\mathbf{J}$.

$$div_r \left[ \mathbf{A}^{ATM}_{B-S} \right] \equiv div_r \left[ \iiint_{all\ space} \frac{d^3 r_s}{4\pi} \frac{\mu_0 \mathbf{J}[t,\mathbf{r}_s]}{\|\mathbf{r}-\mathbf{r}_s\|} \right] = \iiint_{all\ space} \frac{d^3 r_s}{4\pi} \frac{div[\mu_0 \mathbf{J}[t,\mathbf{r}_s]]}{\|\mathbf{r}-\mathbf{r}_s\|} \qquad (\text{III.3})$$

Of course, the Biot-Savart vector potential is not the only one that gives the ATM magnetic field. One could add the gradient of a scalar. For example, one could add the second term on the right hand side of Eq.III.4 and thereby obtain a divergence-free vector potential. The result is ATM's Coulomb gauge vector potential, $\mathbf{A}^{ATM}_C$, which is defined by $div[\mathbf{A}] = 0$.

$$\mathbf{A}^{ATM}_C[t,\mathbf{r}] = \mathbf{A}^{ATM}_{B-S}[t,\mathbf{r}] + \nabla_r \underbrace{\iiint_{all\ space} \frac{d^3 z}{4\pi} \frac{\iiint_{all\ space} \frac{d^3 r_s}{4\pi} \frac{div[\mu_0 \mathbf{J}[t,\mathbf{r}_s]]}{\|\mathbf{r}_s - \mathbf{z}\|}}{\|\mathbf{r}-\mathbf{z}\|}}_{\equiv \Gamma_{B-S \to C}} \qquad (\text{III.4a})$$

$$\mathbf{A}^{ATM}_C[t,\mathbf{r}] = \iiint_{all\ space} \frac{d^3 r_s}{4\pi} \frac{\mu_0 \mathbf{J}[t,\mathbf{r}_s]}{\|\mathbf{r}-\mathbf{r}_s\|} + \nabla_r \iiint_{all\ space} \frac{d^3 z}{4\pi} \frac{\iiint_{all\ space} \frac{d^3 r_s}{4\pi} \frac{div[\mu_0 \mathbf{J}[t,\mathbf{r}_s]]}{\|\mathbf{r}_s - \mathbf{z}\|}}{\|\mathbf{r}-\mathbf{z}\|} \qquad (\text{III.4b})$$

We draw attention to the Coulomb Gauge because it is often used in standard (Maxwell's) electrodynamics. And, we will use it to express the electric field in terms of its sources.

## IV. How the ATM's electric field depends on charges and currents

Until now our focus has been on the magnetic field. But, what of the electric field? Just as with Maxwell's Equations, Faraday's Law, $\mathbf{curl}[\mathbf{E}] = -\partial_t \mathbf{B}$, and the absence of magnetic poles, $div[\mathbf{B}] = 0$, together entail that $\mathbf{curl}[\mathbf{E} + \partial_t \mathbf{A}] = 0$ or $\mathbf{E} = -\nabla \varphi - \partial_t \mathbf{A}$. Thus just as is Maxwell's electrodynamics, ATM is gauge invariant. We choose the Coulomb gauge because, in that gauge, the equation $div[\mathbf{E}] = \rho/\varepsilon_0$ entails the familiar Poisson Equation $-\nabla^2 \varphi = \rho/\varepsilon_0$ having as its solution the familiar Coulomb potential. Thus ATM's electric field is given by Eq.IV.1

$$\mathbf{E}^{ATM} = -\nabla \varphi_C - \partial_t \mathbf{A}_C^{ATM} = -\nabla \varphi_C - \partial_t \mathbf{A}_{B-S}^{ATM} - \partial_t \nabla \Gamma_{B-S \to C} \tag{IV.1}$$

While both the Coulomb scalar potential and the Biot-Savart vector potential are familiar and more or less transparent, the complex expression, denoted by $\Gamma_{B-S \to C}$ is not. Indeed, it appears unfamiliar and difficult to evaluate because of its sixfold integration which is a convolution of two three dimensional integrals. Despite this, the following will show that it is not as bad as it looks. In fact, the following will show it is no worse than the Biot-Savart vector potential.

We wish to simplify $\Gamma_{B-S \to C}$.

$$\Gamma_{B-S \to C} \equiv \iiint_{all\ space} \frac{d^3z}{4\pi} \frac{\iiint_{all\ space} \frac{d^3 r_s}{4\pi} \frac{div[\mu_0 \mathbf{J}[t, \mathbf{r}_s]]}{\|\mathbf{r}_s - \mathbf{z}\|}}{\|\mathbf{r} - \mathbf{z}\|} \tag{IV.2}$$

In particular, we wish to clearly see the relation between the current density at the source point, $\mathbf{r}_s$, and the vector potential at the field point, $\mathbf{r}$. Remarkably, this can actually be arranged. To do so, we re-express the integrand, $div[\mathbf{J}]/\|\mathbf{r}_s - \mathbf{z}\|$, as $div_{r_s}[\mathbf{J}/\|\mathbf{r}_s - \mathbf{z}\|] - \mathbf{J} \cdot \nabla_{r_s}(1/\|\mathbf{r}_s - \mathbf{z}\|)$ and note that the surface integral concomitant with the total divergence is zero because of the current density's bounded support. Thus, we have

$$\Gamma_{B-S \to C}[t, \mathbf{r}] = -\iiint_{all\ space} \frac{d^3 z}{4\pi} \frac{\iiint_{all\ space} \frac{d^3 r_s}{4\pi} \left( \mathbf{J}[t, \mathbf{r}_s] \cdot \nabla_{r_s} \left( \frac{1}{\|\mathbf{r}_s - \mathbf{z}\|} \right) \right)}{\|\mathbf{r} - \mathbf{z}\|} \tag{IV.3}$$

We recast the above by evaluating the gradient with respect to $\mathbf{r}_s$, the source point, and then interchanging the order of $r_s$ and $z$ integrations. The result is shown in Eq.IV.4.

$$\Gamma_{B-S \to C}[t, \mathbf{r}] = -\iiint_{all\ space} \frac{d^3 r_s}{4\pi} \mu_0 \mathbf{J}[t, \mathbf{r}_s] \cdot \left( \iiint_{all\ space} \frac{d^3 z}{4\pi} \left( \frac{1}{\|\mathbf{r} - \mathbf{z}\|} \frac{\mathbf{z} - \mathbf{r}_s}{\|\mathbf{z} - \mathbf{r}_s\|^3} \right) \right) \tag{IV.4}$$

The above presents an integral over $z$ for each source point and each field point. This is a purely geometric problem. If the integral could be evaluated in terms of elementary functions, the result would display what is now implicit, the effect of the current density, **J**, at the source point, $\mathbf{r}_s$, on the value of $\Gamma_{B-S\to C}$ at the field point, **r**. Happily, this can be done.

To do it, we will pick a generic source point $\mathbf{r}_s$, imagine it to be the origin and then integrate over $\mathbf{z} - \mathbf{r}_s$. Having done that for each source point, it will remain only to integrate over all the source points. To prepare the way, we introduce some notation. The difference between field and source points, $\mathbf{r} - \mathbf{r}_s$, will be denoted by $\Delta_{rr_s}$, that is $\Delta_{rr_s} \equiv \mathbf{r} - \mathbf{r}_s$ and the difference between **z** and $\mathbf{r}_s$, will be denoted by $\boldsymbol{\rho}$, that is $\boldsymbol{\rho} \equiv \mathbf{z} - \mathbf{r}_s$. Using this, we rewrite the integral over **z** as follows.

$$\iiint_{all\ space} \frac{d^3z}{4\pi} \left( \frac{1}{\|\mathbf{r}-\mathbf{z}\|} \frac{\mathbf{z}-\mathbf{r}_s}{\|\mathbf{z}-\mathbf{r}_s\|^3} \right) = \iiint_{all\ space} \frac{d^3\rho}{4\pi} \left( \frac{1}{\|\Delta_{rr_s} - \boldsymbol{\rho}\|} \left( \frac{\hat{\boldsymbol{\rho}}}{\|\boldsymbol{\rho}\|^2} \right) \right) \equiv \mathbf{I}[\Delta_{rr_s}] \qquad (IV.5)$$

We will evaluate $\mathbf{I}[\Delta_{rr_s}]$ by using spherical polar coordinates. In particular, we put the origin at $\boldsymbol{\rho} = 0$ and we orient these coordinates so that the line from the origin to the north pole lies on $\Delta_{rr_s}$. Denoting the cosine of the polar angle, $\theta$, by $\chi$ (i.e., $\chi \equiv Cos[\theta]$), we must evaluate

$$I[\Delta_{rr_s}] = \int_0^\infty d\rho \rho^2 \int_{-1}^{+1} d\chi \int_0^{2\pi} \frac{d\varphi}{4\pi} \frac{\rho Sin[\theta]Cos[\varphi]\hat{\mathbf{e}}_1 + \rho Sin[\theta]Sin[\varphi]\hat{\mathbf{e}}_2 + \rho Cos[\theta]\hat{\Delta}_{rr_s}}{\rho^3 \sqrt{\rho^2 + \|\Delta_{rr_s}\|^2 - 2\rho\|\Delta_{rr_s}\|\chi}} \qquad (IV.6)$$

where $\hat{\mathbf{e}}_1$ and $\hat{\mathbf{e}}_2$ are orthogonal to each other and to $\hat{\Delta}_{rr_s}$. Upon integrating over the azimuthal angle, $\varphi$, one finds that the only non-zero contribution is proportional to $\hat{\Delta}_{rr_s}$.

$$\mathbf{I}[\Delta_{rr_s}] = \left( \frac{\hat{\Delta}_{rr_s}}{2} \right) \int_0^\infty d\tilde{\rho} \int_{-1}^{+1} d\chi \frac{\chi}{\sqrt{\tilde{\rho}^2 + 1 - 2\tilde{\rho}\chi}} \qquad (IV.7)$$

This integral can actually be done analytically, though not without an irritation.[7] The form of the integral over $\chi$ depends on whether $\tilde{\rho}$ is more or less than one, as shown below.

$$0 \leq \tilde{\rho} \leq 1 \quad \int_{-1}^{+1} d\chi \frac{\chi}{\sqrt{\tilde{\rho}^2 + 1 - 2\tilde{\rho}\chi}} = \frac{2}{3}\tilde{\rho} \qquad (IV.8a)$$

$$1 \leq \tilde{\rho} \leq \infty \quad \int_{-1}^{+1} d\chi \frac{\chi}{\sqrt{\tilde{\rho}^2 + 1 - 2\tilde{\rho}\chi}} = \frac{2}{3}\tilde{\rho}^{-2} \qquad (IV.8b)$$

Using both results to integrate $\tilde{\rho}$ from 0 to $\infty$, one finds that the numerical value of the multiple integral is 1.

$$\int_0^\infty d\tilde{\rho} \int_{-1}^{+1} d\chi \frac{\chi}{\sqrt{\tilde{\rho}^2 + 1 - 2\tilde{\rho}\chi}} = 1 \qquad (IV.9)$$

And so, one sees that that the integral, $\mathbf{I}[\Delta_{rr_s}]$, is simply half the unit vector pointing from $\mathbf{r}_s$ to $\mathbf{r}$.

$$\mathbf{I}[\Delta_{rr_s}] = \frac{1}{2}\hat{\Delta}_{rr_s} \tag{IV.10}$$

Thus $\Gamma_{B-S \to C}$, the quantity that manifests the time dependence of the charge density, can be written as shown next

$$\Gamma_{B-S \to C}[t,\mathbf{r}] = -\iiint_{all\ space} \frac{d^3 r_s}{4\pi}\left(\mu_0 \mathbf{J}[t,\mathbf{r}_s] \cdot \frac{1}{2}\hat{\Delta}_{rr_s}\right) \tag{IV.11}$$

and its gradient appears below.

$$\nabla_r \Gamma_{B-S \to C}[t,\mathbf{r}] = -\frac{1}{2}\iiint_{all\ space} \frac{d^3 r_s}{4\pi}\left(\frac{\mu_0 \mathbf{J}[t,\mathbf{r}_s]}{\|\mathbf{r}-\mathbf{r}_s\|} - \frac{(\mu_0 \mathbf{J}[t,\mathbf{r}_s]\cdot\hat{\Delta}_{rr_s})\hat{\Delta}_{rr_s}}{\|\mathbf{r}-\mathbf{r}_s\|}\right) \tag{IV.12}$$

And so the coulomb gauge vector potential is as shown in Eq.IV.13

$$\mathbf{A}_C^{ATM} = \mathbf{A}_{B-S}^{ATM} + \nabla \Gamma_{B-S \to C} \tag{IV.13a}$$

$$\mathbf{A}_C^{ATM} = \frac{1}{2}\iiint_{all\ space} \frac{d^3 r_s}{4\pi}\left(\frac{\mu_0 \mathbf{J}[t,\mathbf{r}_s]}{\|\mathbf{r}-\mathbf{r}_s\|} + \frac{(\mu_0 \mathbf{J}[t,\mathbf{r}_s]\cdot\hat{\Delta}_{rr_s})\hat{\Delta}_{rr_s}}{\|\mathbf{r}-\mathbf{r}_s\|}\right) \tag{IV.13b}$$

Here, it may be helpful to recall where we started, the relation between the electric field and the potentials and thus the relation between the electric field and its sources.

$$\mathbf{E}^{ATM} = -\nabla \varphi_C - \partial_t \mathbf{A}_C^{ATM} = -\nabla \varphi_C - \partial_t \mathbf{A}_{B-S}^{ATM} - \partial_t \nabla \Gamma_{B-S \to C} \tag{IV.1}$$

Note that we have found that while the magnetic field is given by its Biot-Savart vector potential, this vector potential is not sufficient to manifest the electric field induced by a changing magnetic field when the charge density is also changing. As already stated, the changing charge density is manifest in $\Gamma_{B-S \to C}$.

## V. Summary and conclusions

A few points deserve emphasis. The first is straightforward. In the ATM, electric field and magnetic field are determined by charge and current densities, and their time derivatives at the same time as the field is measured. In short, the ATM propagates influence instantaneously. Second, unlike magneto-quasistatic approximation, the ATM enables local charge conservation while including the possibility of a time dependent charge density. Thus the ATM may approximate Maxwell's theory when and where retardation may be neglected. Third, The ATM's scalar and vector potentials depend on their sources charges and currents in the same way that Darwin's Approximation of the Coulomb gauge solutions of Maxwell's Equations. This is remarkable. Darwin got his approximation by assuming his charge carriers were classical point particles interacting with the field and then expanding that system's Lagrangian in inverse powers of the speed of light. The ATM is macroscopic theory that arrives at the same dependence by starting with an alternative to Maxwell's Displacement Current. Fourth, the Coulomb gauge vector potential of ATM does not appear qualitatively more difficult to evaluate that the Biot-Savart vector potential. Thus ATM provides a comprehensive (including both electro and magneto quasistatics) and tractable approximation to Maxwell's equations.